# Nickel Doping Unlocks Ambient-Condition Photostability in Individual Cesium Lead Bromide Perovskite Quantum Dots

*Jehyeok Ryu[1,3,‡], Victor Krivenkov[2,3,‡], Adam Olejniczak[2], Mikel Arruabarrena[1,2], Jozef Janovec[1,2,3],Sebastien E. Hadjadj[2], Maxim Ilyn[2], Aritz Leonardo[1,4], Virginia Martínez-Martínez[5], Andres Ayuela[1,2], Alexey Y. Nikitin[1,6*], Yury Rakovich[1,2,3,6*]*

**Author Address**

[1]Donostia International Physics Center (DIPC), Donostia-San Sebastián 20018, Spain.

[2]Centro de Física de Materiales (CFM-MPC), Donostia - San Sebastián, 20018, Spain

[3]Polymers and Materials: Physics, Chemistry and Technology, Chemistry Faculty, University of the Basque Country (UPV/EHU), Donostia—San Sebastián 20018, Spain

[4]EHU Quantum Center, University of the Basque Country UPV/EHU, 48940 Leioa, Spain

[5]Departamento de Química Física, Universidad del País Vasco, UPV/EHU, Apartado 644, Bilbao, 48080 Spain

[6]IKERBASQUE, Basque Foundation for Science, Bilbao 48013, Spain

[‡]J.R. and V.K. contributed equally to this work.

**Abstract**

Developing efficient single-photon sources is fundamental to advancing photonic quantum technologies. In particular, achieving scalable, cost-effective, stable, and high-purity single-photon emission at ambient conditions is paramount for free-space quantum communication. However, fulfilling all the requirements simultaneously under ambient conditions has remained a significant challenge. Here we present the scalable, cost-effective ambient condition synthesis of nickel doped $CsPbBr_3$ perovskite quantum dots (NPQDs) using a modified ligand-assisted reprecipitation method. The resulting individual NPQDs demonstrate remarkable photostability, sustaining their performance for over 10 minutes under ambient conditions, and exhibit exceptional single-photon purity (>99%) with a narrow emission linewidth (~70 meV). The remarkable photostability could be attributed to the spatial localization of the exciton by Ni atoms on the surface of the nanocrystal, reducing its interaction with the environment. Our results demonstrated that NPQDs with outstanding combinations of quantum emitting properties can be both synthesized and operated at ambient conditions.

Modern quantum technologies are rapidly evolving, with many of them relying on photon-based qubits at their core. Quantum emitters (QEs) capable of generating individual single-photons are crucial for enabling these flying qubits, serving as the fundamental building blocks for secure quantum communication, as well as quantum metrology, and imaging[1]. In some quantum key distribution communication protocols like BB84, security relies on high single-photon purity, ensuring that nearly every qubit is encoded to exactly one photon, greatly reducing the risk of multi-photon events that an eavesdropper could exploit[2]. Similarly, in quantum imaging techniques such as low-light imaging[3], the reliability that each detection event corresponds to a single-photon leads to higher image contrast and resolution by reducing false coincidences and background noise. Therefore, high single-photon purity, superior operational stability at ambient conditions are key characteristics required for QEs to meet the demands of the implementation of scalable, cost-effective quantum technologies for real-world use.

Semiconductor quantum dots (QDs) have emerged as promising and efficient quantum emitters[2]. Notably, epitaxial QDs are already well-established as sources of indistinguishable single-photons at cryogenic temperature[4]. Nonetheless, practical applications such as free-space quantum communication demand quantum light emission at room temperature or even under ambient conditions[5], where epitaxial QDs face challenges in maintaining the single-photon purity and brightness[6]. In contrast, colloidal QDs can operate as single-photon emitters with the required characteristics even at room temperature. Among them, colloidal lead halide perovskite quantum dots (PQDs) have been intensively explored as QEs at both room and cryogenic temperatures[7,8]. These PQDs offer near-unity photoluminescence quantum yield (PLQY), high single-photon purity, and considerably longer coherence times compared to other types of colloidal QDs. Furthermore, PQDs exhibit shortened lifetimes[9] and bright emissive exciton triplet states at low temperature, attributed to inhibited one phonon-assisted exciton transitions to dark singlet ground state[10,11], contrary to conventional colloidal QDs. As recently shown by Kaplan et al.[8], $CsPbBr_3$ PQDs can produce indistinguishable single-photons as verified by Hong–Ou–Mandel interference. Nevertheless, achieving sufficiently high quality stable PQDs remains a critical step for exploring single-photon emission at the individual QD level. Ligand-assisted reprecipitation (LARP) method provides a scalable, cost-effective synthesis of PQD under ambient conditions, but often produces quantum dots of lower quality with broad size distribution–suitable primarily for optoelectronic applications like solar cells and light emitting diodes. In contrast, hot-injection synthesis typically produces superior-quality monodisperse PQDs but requires elevated temperatures (T: 440−550 K), inert atmosphere, increasing the complexity of the synthetic procedure[12]. Furthermore, PQDs are structurally labile at the individual nanocrystal level, and prone to degrade under illumination as excess photoexcited charges on the surface react with ambient moisture or oxygen[13], limiting their practical applicability as single-photon sources operating at real-world conditions[14]. To reduce the light-induced surface degradation, strategies to isolate PQDs from moisture and oxygen have been widely explored by encapsulating with polymer[15] under inert gas[7] or by engineering ligands for improved passivation[16]. Zhu et al.[7] demonstrated that $CsPbBr_3$ PQDs encapsulated in a polystyrene matrix layer exhibited photostability for over 80 seconds in an inert environment, yet the relatively high value of their second-order cross-correlation function $g^{(2)}(0) \sim 0.27$ indicated non-pure single-photon behavior. Without inert conditions, the PQDs quickly degraded, which was marked by a strong blueshift in their emission spectrum with a further drop in the intensity to zero value. More recently, Morad et al. demonstrated that effective ligand engineering can extend the stable emission of $CsPbBr_3$ PQDs to 3 minutes in an inert environment[16]. Despite these advancements, achieving stable $CsPbBr_3$ PQDs that can operate reliably as single-photon emitters under ambient conditions still remains a major challenge.

An alternative promising strategy for enhancing the stability of PQDs is their doping by transition metal ions[17]. In particular, doping of $CsPbBr_3$ and $CsPbI_3$ PQDs, synthesized by hot-injection method, with $Ni^{2+}$ ions has been shown to boost PLQY and environmental stability of PQD ensembles in solutions or dense films[18–23]. However, for the design of quantum emitters (QEs), such ensemble architectures are unsuitable, as excitation and photon collection must occur from the same individual nanocrystal without interaction with neighboring ones. For quantum-emitting applications, the relevant performance metrics of individual PQDs differ substantially from those of PQD ensembles for LED applications: high single-photon purity, narrow linewidth, short radiative lifetime, and long-term spectral stability at the single QE level are required. In single QEs, spectral blueshift and size reduction are among the primary mechanisms of photodegradation under ambient conditions, which critically limit their applicability[7,15,24,25], while in dense films or colloidal solutions, the predominant stability concern is the loss of ensemble PL intensity[26]. Thus, despite the demonstrated benefits of $Ni^{2+}$ doping for ensemble PQDs, its specific impact on the quantum-emitting properties of individual PQDs remains largely unexplored.

In this study, we address this critical gap by devising a synthesis protocol for Ni-doped $CsPbBr_3$ PQDs (NPQDs). By incorporating $Ni^{2+}$ ions during the glovebox-free LARP synthesis at ambient conditions (room temperature, 70% relative humidity), we obtained highly stable NPQDs solutions with PLQY ~86%. We demonstrated their superior performance by rigorously testing their performance as QEs by evaluating their photostability, photoluminescence (PL) linewidth, PL lifetime, and single-photon purity under ambient conditions. Notably, this work represents perovskite QDs synthesized entirely under ambient conditions that function as high-purity single-photon sources in the same environment, paving the way for more accessible and scalable quantum technologies.

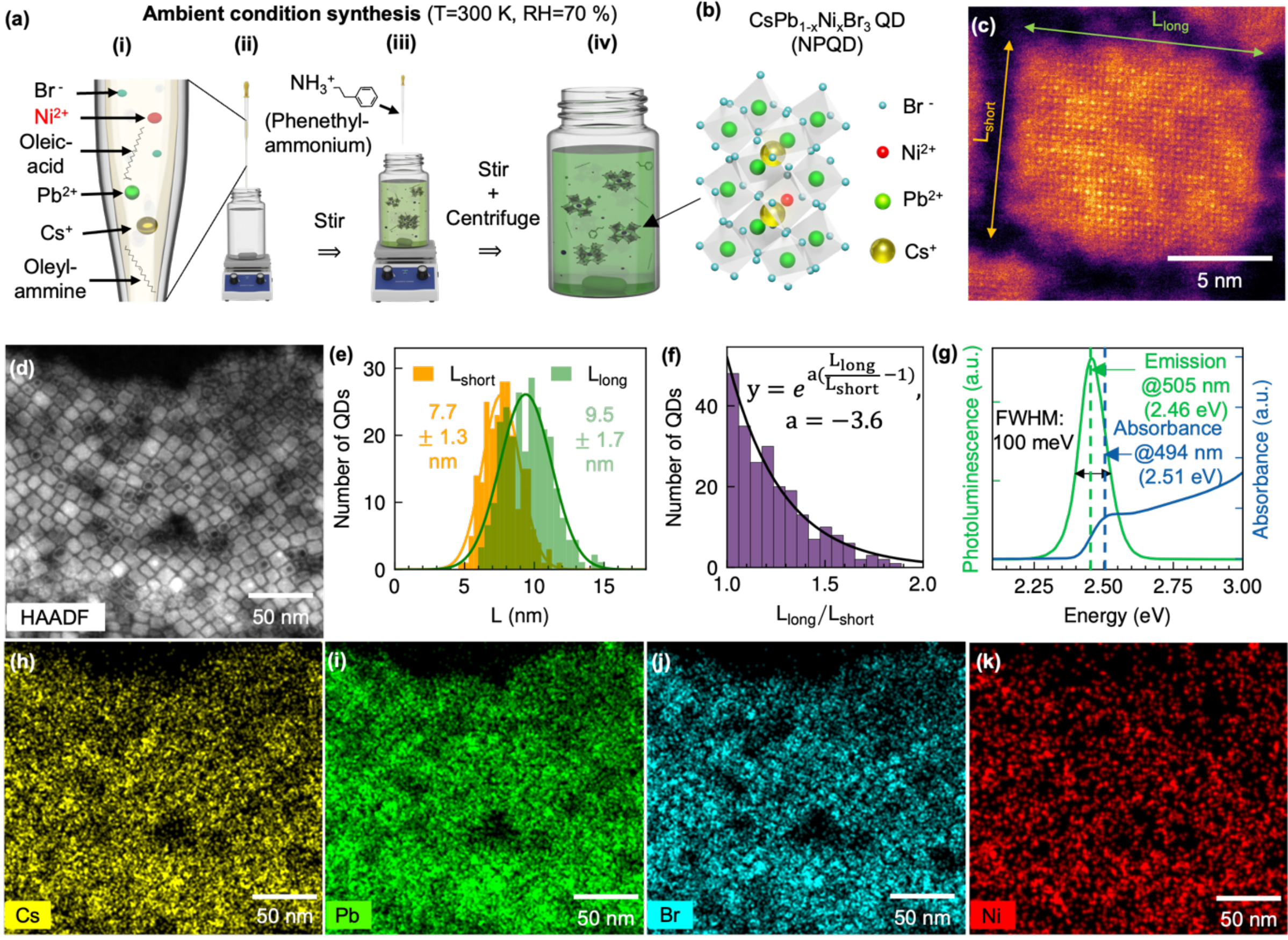

Figure 1. Synthesis and characterization of Ni-doped $CsPbBr_3$ PQDs (NPQDs). (a) Schematic illustration of ligand assisted reprecipitation synthesis under ambient conditions. (b) Schematic crystal structure of NPQD (c, d) HAADF-STEM of single NPQD (c) and NPQDs (d). (e) The distribution of shorter and longer edge lengths ($L_{short}$, and $L_{long}$) of each NPQDs and (f) the histogram of the ratio $L_{long}/L_{short}$. (g) Absorbance and PL spectra of NPQD solution. (h-i) The EDS elemental maps of Cs L lines (h), Pb L lines (i), Br K lines (j), and Ni k line (k) (the energies are selected according to Fig. S1) in the same scale and area as the HAADF-STEM image (d).

We synthesized NPQDs by modifying the LARP[27] method under ambient conditions, as schematically illustrated in Fig. 1a. The process involves four key steps: (i) preparation of the seed solution of CsBr, $PbBr_2$, and $NiBr_2$ (molar ratio 1:1:0.5)[18] in dimethylformamide (DMF) with oleic acid and oleylamine, followed by sonication for 1 hour to effectively dissolve $NiBr_2$; (ii) inducing nucleation and formation of nanocrystals by injecting the seed solution into toluene; (iii) ligand exchange from highly dynamic, long-chain ligand (oleylamine) to the more stable, low-steric hindrance, short-chain ligand[28,29] (phenethylammonium, PEA) to enhance nanocrystal stability; and (iv) purification via centrifugation to obtain monodisperse nanocrystals (see Methods for details). The entire synthesis was performed at room temperature and 70% relative humidity, demonstrating the viability of an ambient-condition synthesis approach. Figure 1b depicts a 3D illustration of the expected crystal structure of a NPQD, where $Ni^{2+}$ ions replace $Pb^{2+}$ ions in the crystal lattice, forming octahedral coordination with halide ions[20].

High-angle annular dark field scanning transmission electron microscopy (HAADF-STEM) images (Fig. 1c, d) reveal that the synthesized NPQDs retained a cuboid morphology. Notice that dark spots observed on the NPQDs in Fig. 1d do not reflect any degradation in their quality as the damages result from the electron beam. To quantify any Ni-induced shape change, we measured the shorter and longer edge lengths ($L_{short}$ and $L_{long}$ in Fig. 1c) of each particle in the HAADF-STEM image (Fig. 1d) and plotted the distributions in Fig. 1e. The distributions of edge lengths are narrow, with mean values of $7.7 \pm 1.3$ nm for $L_{short}$ and $9.5 \pm 1.7$ nm for $L_{long}$, indicating that the synthesized NPQDs are monodisperse. A histogram of the aspect ratio $L_{long}/L_{short}$ (Fig. 1f) shows that cubic particles with aspect ratio near-unity dominate the population and that the number of higher aspect ratio cuboids falls off exponentially. The mean aspect ratio is 1.2, which matches the value reported for pristine $CsPbBr_3$ nanocrystals[30]. These results demonstrate that the introduction of Ni does not significantly alter the particle morphology.

To confirm the incorporation of Ni into the NPQDs, we conducted energy-dispersive X-ray spectroscopy (EDS) spectrum (Fig. S1) and elemental maps (Fig. 1h-k, Fig. S3). Comparison of the HAADF-STEM image (Fig. 1d) with the EDS maps shows that Cs, Pb, Br, and Ni signals emanate from the NPQDs themselves. XPS measurements revealed that the sample contained Br, Pb, Cs, and Ni in the correct configurations (Figure S7). The presence of Ni in the nanocrystals is verified with the Ni 2p peaks (Figure S7a). Despite a high molar concentration of Ni during the synthesis, only a small fraction of Ni is incorporated into the nanocrystal, as evidenced by the small molar ratio of Ni/Pb ~ 0.39% from an independent ICP-AES measurement (Fig. S2). This low incorporation level indicates that $Ni^{2+}$ ions tend to be doped rather than forming an alloy. Additionally, selected area electron diffraction (SAED) patterns (Fig. S4) show dominant rings corresponding to the (100), (110), and (200) crystal planes of $CsPbBr_3$ QDs[31], further supporting that the Ni inside the crystal improves crystallinity, which would reduce structural lability[18].

Optical characterization further evidenced the impact of Ni doping on the optical properties of the NPQD solution. Absorbance and PL spectra of NPQD solution (Fig. 1g) show a first exciton peak at 2.51 eV and an emission peak at 2.46 eV with a linewidth of 100 meV. The NPQD solution exhibits a Stokes shift of about 50 meV, consistent with previously reported values for PQD solutions[32]. In addition, the emission peak energy of the NPQD solution (Fig. S5a) is only slightly blue-shifted, and the PL lifetime of the NPQD solution (Fig. S5b) is only modestly shortened compared to the previously reported $CsPbBr_3$ solution. This result implies that the exciton-phonon interaction and the energy levels remained similar despite Ni incorporation.

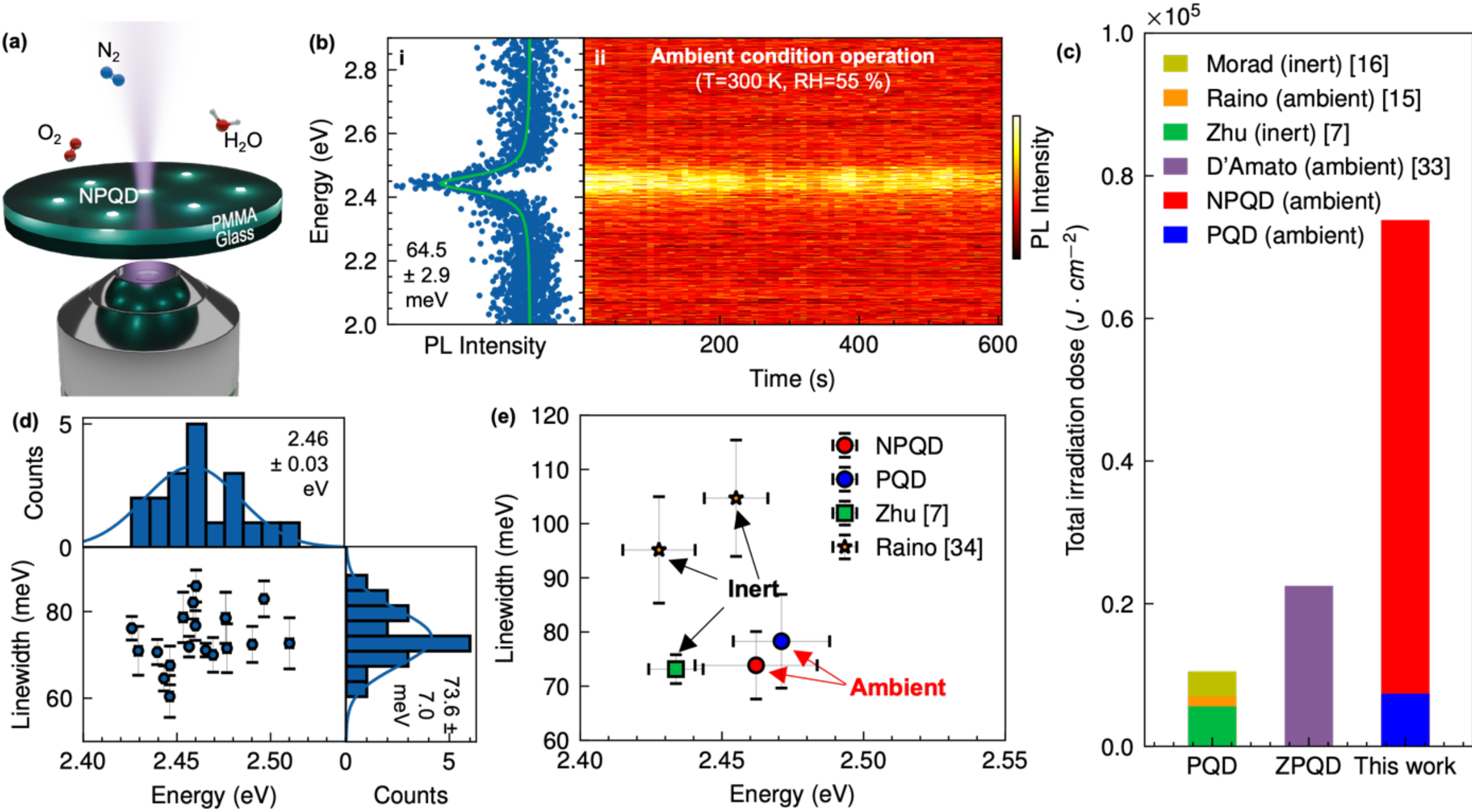

Figure 2. Photostability and PL linewidths of individual NPQDs in PMMA thin film matrix measured under ambient conditions. (a) Schematic illustration of NPQDs embedded in thin PMMA film spin-coated onto a cover glass. (b) PL spectrum of a representative single NPQD (i) and sequential PL spectrum temporal trace measured for 10 mins (ii). (c) Comparison between the studied NPQDs and previously reported values for conventional $CsPbBr_3$ QDs [7,15,16,33] of the total laser irradiation dose before the spectral shift occurred. (d) PL spectrum linewidths distribution of individual NPQDs and (e) comparison with previously reported values for $CsPbBr_3$ PQDs[7,34]

High-purity single-photon emission is a fundamental requirement for quantum photonic technologies, enabling secure quantum communication. However, maintaining photostable high single-photon purity remains a challenge. To assess the emission photostability and quantum emitting characteristics of NPQDs, we prepared thin NPQD-PMMA films with 45 nm thickness (estimated according to Ref.[35]), by spin-coating diluted NPQD-PMMA solutions onto glass substrates. PL measurements of individual NPQDs were conducted under ambient conditions (T = 300 K, RH = 55%), as illustrated in Fig. 2a.

To assess emission photostability, individual NPQDs were excited at 405 nm using a continuous-wave laser source (123 W·cm$^{-2}$). We recorded the PL spectral trace of single NPQDs for 10 minutes with 10-s photon accumulation time for each measured PL spectrum. This allowed us to monitor the real-time evolution of PL spectra and assess the PL spectral stability. Remarkably, NPQDs displayed narrow linewidths down 64.5 meV and remained highly stable for up to 10 minutes without noticeable spectral shifts and significant intensity loss (Fig. 2bii). To compare photostability, we plotted total irradiance dose (averaged power density × irradiated time, J·cm$^{-2}$), representing total energy exposure. Previous studies have reported that undoped 10 nm $CsPbBr_3$ PQDs embedded in polystyrene films sustained emission for up to only 100 s at 70 W·cm$^{-2}$ under ambient condition[15], corresponding to 7000 J·cm$^{-2}$ irradiation dose (orange bar in Fig. 2c), thus demonstrating a low photostability. In an inert atmosphere, emission improved but remained below 200 s as indicated by green (5600 J·cm$^{-2}$) and yellow bar (10500 J·cm$^{-2}$) in Fig. 2c[7,16]. Zn-doped $CsPbBr_3$ PQDs have been reported to maintain emission for 60 minutes at 6.5 W·cm$^{-2}$, corresponding to 22500 J·cm$^{-2}$ of irradiation dose, demonstrating improved photostability via metal doping[33] (purple bar in Fig. 2c). Notably, our NPQDs exhibit an irradiation dose of 73800 J·cm$^{-2}$, indicating a dramatic extension of photostability (red bar in Fig. 2c), suggesting that Ni incorporation effectively suppresses common degradation pathways, and highlighting their potential for practical single-photon applications. The corresponding comparison of photostability data reported in the literature for individual $CsPbBr_3$ PQDs could be found in Table S1 in the Supporting Information file.

Furthermore, we investigate the emission linewidths of individual NPQDs, as it is a direct indicator of dephasing. Linewidth broadening is mainly driven by exciton-phonon coupling and interactions with the environment. To assess linewidth distribution, we plotted the first 10-second accumulated PL spectra of 20 individual NPQDs (Fig. 2d). The statistical distribution of linewidths follows a Gaussian distribution with a mean value of 73.6 ± 7.0 meV, and emission peaks centered at 2.46 ± 0.03 eV. Remarkably, the linewidth of NPQDs is similar to that of high-quality undoped 20 nm $CsPbBr_3$ PQDs, which highlights the excellent optical performance of our QDs at room temperature[8]. Furthermore, comparing linewidths as a function of emission energy reveals that NPQDs operating under ambient conditions (red circle in Fig. 2e) have linewidths comparable to the best values reported for $CsPbBr_3$ PQDs operated under an inert environment (green square in Fig. 2e)[7].

We hypothesize that the enhanced photostability and narrower linewidth observed in individual NPQDs arise from the incorporation of Ni into the nanocrystal, which reduces exciton–environment interactions. To support this hypothesis, we performed density functional theory (DFT) calculations of the crystal formation energies for both pristine and Ni-doped bulk $CsPbBr_3$ to investigate the most favorable position of the Ni atom within the crystal. The calculations indicate that both pristine and Ni-doped $CsPbBr_3$ are thermodynamically stable, confirming that Ni atoms can indeed be incorporated in the crystal (Fig. S8a). Nevertheless, the formation energy of the Ni-doped $CsPbBr_3$ is noticeably higher than that of pristine $CsPbBr_3$ (see Supplementary Notes (SN) in the Supporting Information for the details), implying that Ni atoms energetically tend to migrate to the surface (Fig. S8b).

To understand how the Ni atoms affect the band-edge exciton, we studied the charge density of the latter for both pristine and Ni-doped bulk $CsPbBr_3$ (Fig. S9c, d, see SN for the details). In pristine $CsPbBr_3$, the band-edge exciton is delocalized across the lattice, predominantly around Pb atoms, implying a high probability of interaction between the exciton and the environment (Fig. S9c). Even though the exciton in the Ni-doped $CsPbBr_3$ is delocalized across the lattice around Pb atoms, it is unconfined to Ni atoms (Fig. S9d). Hence, we anticipate that the exciton in the Ni-doped $CsPbBr_3$ nanocrystal is unconfined to Ni atoms at the surface, reducing the interaction with the environment. Furthermore, we calculated the defect formation energies for both pristine and Ni-doped $CsPbBr_3$ to evaluate whether Ni incorporation reduces defect formation (Fig. S11). The DFT results show that the formation energy of the bromide vacancies is comparable, or even slightly lower, in Ni-doped $CsPbBr_3$ compared to pristine $CsPbBr_3$. This indicates that Ni doping does not lead to reduced defect formation. Therefore, we conclude that the enhanced photostability observed in Ni-doped $CsPbBr_3$ is not related to the defect passivation. In brief, we speculate that our DFT calculations provide a hint that the spatial localization of the band edge exciton in the presence of Ni atoms can be a path to improve photostability by reducing interaction with the environment.

Our experimental and theoretical analyses demonstrate that Ni-doping not only enhances the operational stability of NPQDs under ambient conditions but also maintains their remarkable optical quality, making them promising for practical quantum photonic applications.

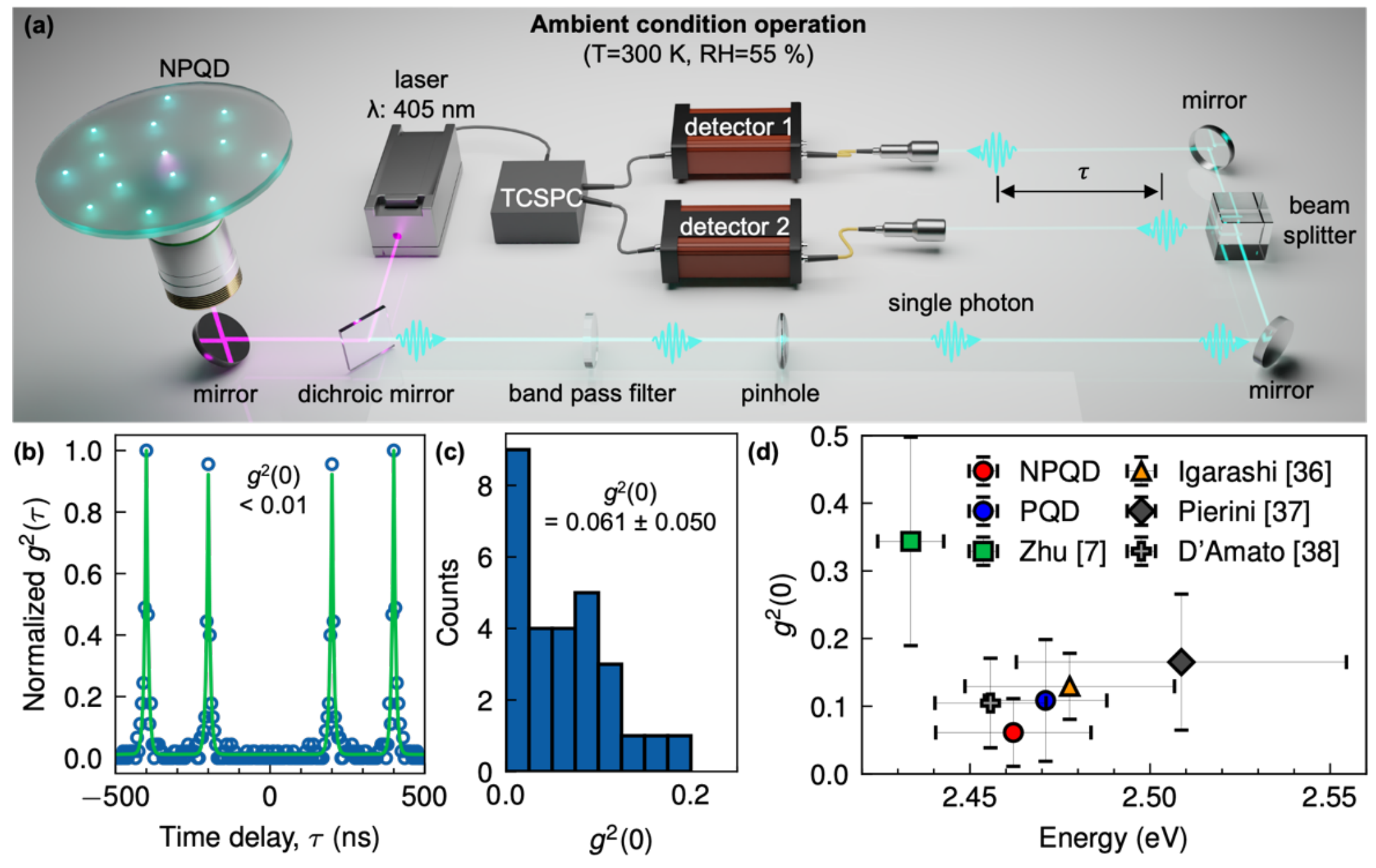

Figure 3. Single-photon purity of individual NPQD. (a) Schematic illustration of Hanbury Brown Twiss setup for verification of single-photon purity ($g^{(2)}(\tau)$ measurements) (b) $g^{(2)}(\tau)$ of the representative NPQD. (c) Statistics of $g^{(2)}(0)$ values of NPQDs and (d) their comparison with previously reported values for $CsPbBr_3$ PQDs[7,36–38].

To assess the single-photon emission characteristics of NPQDs, we performed Hanbury-Brown and Twiss (HBT) measurements, which provide insights into photon statistics from individual NPQDs (Fig. 3a). Using a pulsed laser at 405 nm with a 5 MHz repetition rate, ~200 ps pulse duration, and 31 W $cm^{-2}$ average intensity, we excited NPQDs embedded in PMMA on glass substrates. Due to demonstrated high photostability over 200 s for all individual NPQDs (Fig. S6), we could conclude that the $g^{(2)}$ measurements, performed only during 100 s, were unaffected by the photoinduced spectral blue-shift, which could be attributed to the size shrinkage and thus the change of the single-photon purity properties, in contrast to previously reported results for $CsPbBr_3$ PQDs[7,13].

The second-order correlation function, $g^{(2)}(\tau)$, was reconstructed from a time-tagged time-resolved photon arrival data array without any background noise subtraction, for 30 individual NPQDs. Notably, our best-performing individual NPQDs exhibited $g^{(2)}(0) < 0.01$ under ambient conditions without any temporal or spectral filtering (Fig. 3b), indicating high-purity single-photon emission from synthesized nanocrystals. Among the 30 individual NPQDs tested, all exhibited $g^{(2)}(0) < 0.2$, with over 70% displaying values below 0.1, with a mean value $g^{(2)}(0)$ = 0.061 ± 0.050 (Fig. 3c). This high level of single-photon purity rivals even epitaxial quantum dots, which often require cryogenic conditions for comparable performance. These findings underscore the strong potential of NPQDs as a highly promising platform for practical, scalable quantum photonics. In Figure 3d, we compare the mean $g^{(2)}(0)$ value measured from our NPQDs with previously reported values measured at room temperature [7,36–38]. Notably, our values surpass previously reported values of $CsPbBr_3$ PQDs. Specifically, photostable individual 10 nm $CsPbBr_3$ PQDs in an inert atmosphere (green square in Fig. 3d) have a rather low single-photon purity ($g^{(2)}(0)$ > 0.3)[7], indicating insufficient suppression of multiexciton generation at this size. Besides, the three points (grey, orange, black points) in Fig. 3d show relatively low $g^{(2)}(0)$ values (0.1~0.2), however details about the operating atmosphere condition and photostability were not explicitly provided. It is known that relatively low $g^{(2)}(0)$ values can result from increased confinement due to size shrinkage by photodegradation[7]. In contrast, our individual NPQDs exhibit high single-photon purity while maintaining their photostability, demonstrating their feasibility as efficient single-photon sources. The corresponding comparison of single-photon purity data reported in the literature for individual $CsPbBr_3$ PQDs could be found in Table S1 in the Supporting Information file.

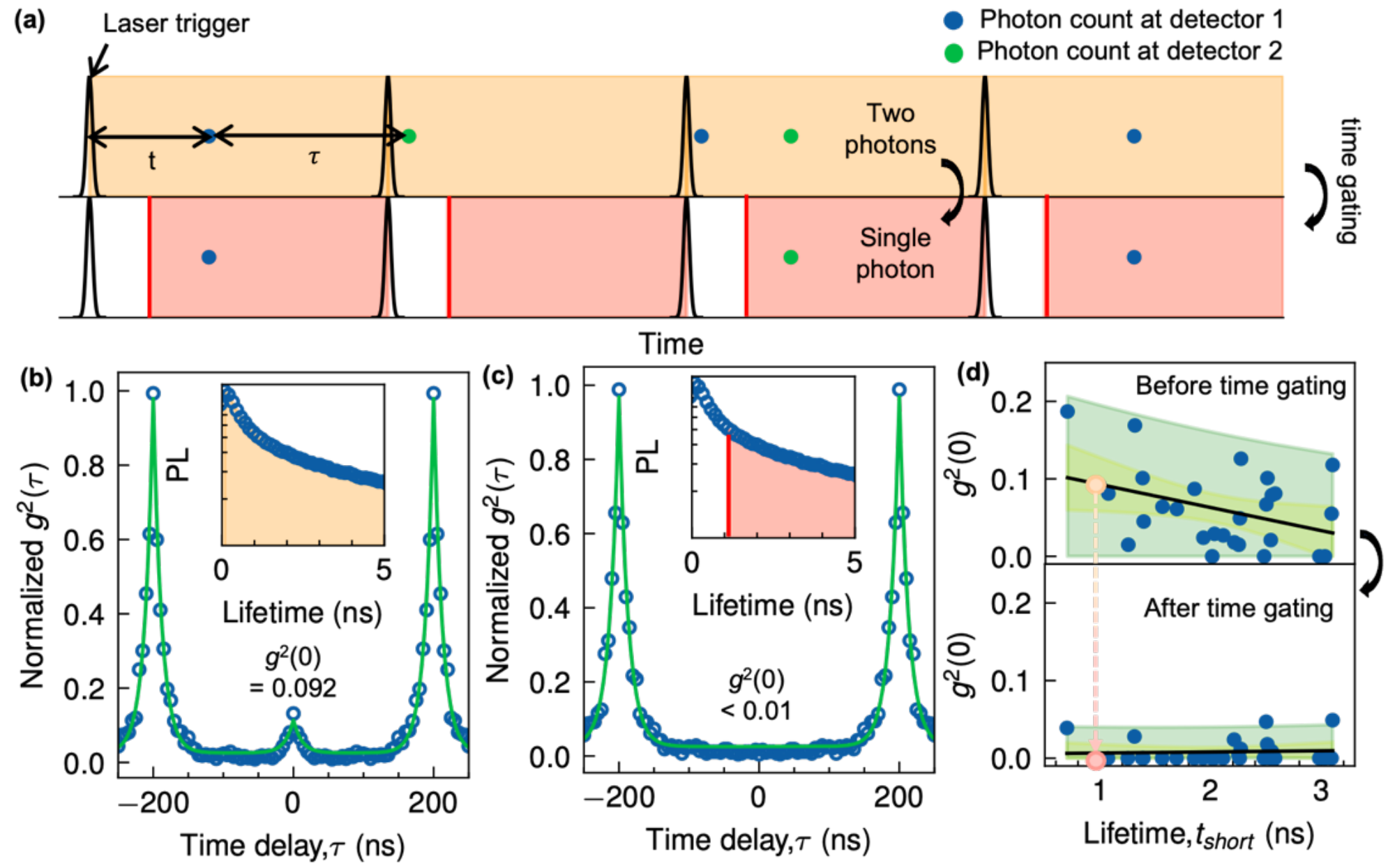

Figure 4. Time-gated single-photon purity of individual NPQDs (a) Schematic illustration of the time-gating process. (b) Normalized $g^{(2)}(\tau)$ of the representative individual NPQD before time-gating, which corresponds to the apricot circle in (d); (c) the same after time-gating, which corresponds to the red circle in (d). (d) $g^{(2)}(\tau)$ values statistics of individual NPQDs before (top) and after (bottom) time-gating.

We hypothesize that the non-zero $g^{(2)}(0)$ values in our NPQDs (indicating the non-ideal single-photon emission) arise from the multi-photon emission from NPQDs themselves and not from the environment. Indeed, biexciton-exciton cascades can lead to two-photon emission in a quantum dot [39]. To verify the origin of non-zero $g^{(2)}(0)$ values in some of the NPQDs, we performed time-gating analysis of the emission signal (Fig. 4a). In this technique, multi-photon emissions that came in a certain time range after the excitation pulse are excluded. As illustrated in the top part of Fig. 4a, two photons can be registered after a laser pulse. However, with time gating (bottom of Fig. 4a), certain time windows from each laser pulse are excluded, maintaining pure single-photon statistics. In $CsPbBr_3$ PQDs, the biexciton-exciton recombination occurs typically on the order of tens of picoseconds[39], which is significantly faster than exciton-ground recombination decay, occurring on the order of a few nanoseconds. Thus, if the main reason of the non-zero $g^{(2)}(0)$ level is the biexciton emission, then the time gating technique should reduce this level to near zero. Indeed, by applying a 1 ns time gate which selectively excludes early-arriving photons associated with biexciton recombination (inset in Fig. 4c), we were able to effectively isolate single-photon emission so that the $g^{(2)}(0)$ value for a representative NPQD was reduced from approximately 0.092 (Fig. 4b) to less than 0.01 (Fig. 4c). To confirm the improvement in single-photon purity for the whole set of NPQDs after time-gating, we plotted $g^{(2)}(0)$ values against $t_{short}$, where $t_{short}$ is the short-lifetime component obtained from a typical biexponential lifetime fit (Fig. 4d). Before applying time-gating, the $g^{(2)}(0)$ values tended to increase at shorter $t_{short}$ due to increased contributions from photons originating from biexcitons (Fig. 4d, top). However, a significant percentage of NPQDs exhibited $g^{(2)}(0)$ values reduced to less than 0.01 after time gating (Fig. 4d, bottom), supporting our hypothesis that the main contribution to non-zero $g^{(2)}(0)$ value is the emission from the biexciton-exciton transition. These results demonstrate that Ni doping significantly enhances the single-photon purity of $CsPbBr_3$ PQDs, achieving high single-photon purity while maintaining exceptional photostability under ambient conditions. The combination of high single-photon purity, extended emission stability, and the ambient-compatible synthesis positions NPQDs as promising candidates for practical quantum communication and photonic circuit applications.

Thus, we have overcome significant barriers in the development of high-performance single-photon quantum sources by demonstrating a straightforward method to synthesize high-quality perovskite quantum dots entirely under ambient conditions. The introduction of Ni doping has not only enhanced the photostability of the $CsPbBr_3$ PQDs but also enabled single-photon emission with exceptional purity exceeding 99%. This success allows the development of stable and efficient single-photon sources based on NPQDs integrated into real devices without the need for a controlled environment. The cost-effectiveness and simplicity of the synthesis process open up opportunities for a broad range of applications, including quantum communications and cryptography, photonic circuits, and optoelectronic devices where stability under ambient conditions is paramount. Moreover, the demonstrated ability to fabricate high-quality NPQDs in thin film lays the groundwork for coupling them with photonic cavities and exploring their integration into quantum photonic circuits.

**Supporting Information:** EDS spectra and image map, ICP-AES measurements, SAED, and photostability spectra of the Ni-doped $CsPbBr_3$ QDs, DFT calculation

**Corresponding Author**

Alexey Y. Nikitin (alexey@dipc.org), Yury Rakovich (yury.rakovich@ehu.eus)

**Acknowledgement**

The study was funded by the Department of Science, Universities and Innovation of the Basque Government (grants no. IT1526-22, PIBA-2023-1-0007, IT1639-22, IT-1569-22, PIBA-2024-1-0011) and the IKUR Strategy; by the Spanish Ministry of Science and Innovation (grants no. PID2022-141017OB-I00, TED2021-129457B-I00, PID2023-146442NB-I00, PID2023-147676NB-I00, PID2022-139230NB-I00, TED2021-132074B-C32, PID2022-138750NB-C22, TED2021-130292B-C42). A.A. acknowledges support from the European Commission through the NaturSea-PV project (GA 101084348). Y.R. and A.O. acknowledge support from the ONRG (Award No. N62909-22-1-2031). This research was conducted within the framework of the Transnational Common Laboratories (LTC) Aquitaine-Euskadi Network in Green Concrete and Cement-based Materials and TRANS-LIGHT. Computations for this research were carried out using DIPC supercomputer resources.

**Data Availability Statement**

The data that support the findings of this study are available from the corresponding author upon reasonable request.

**Conflicts of Interest**

The authors declare no conflict of interest.

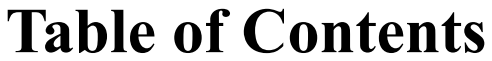

## Table of Contents

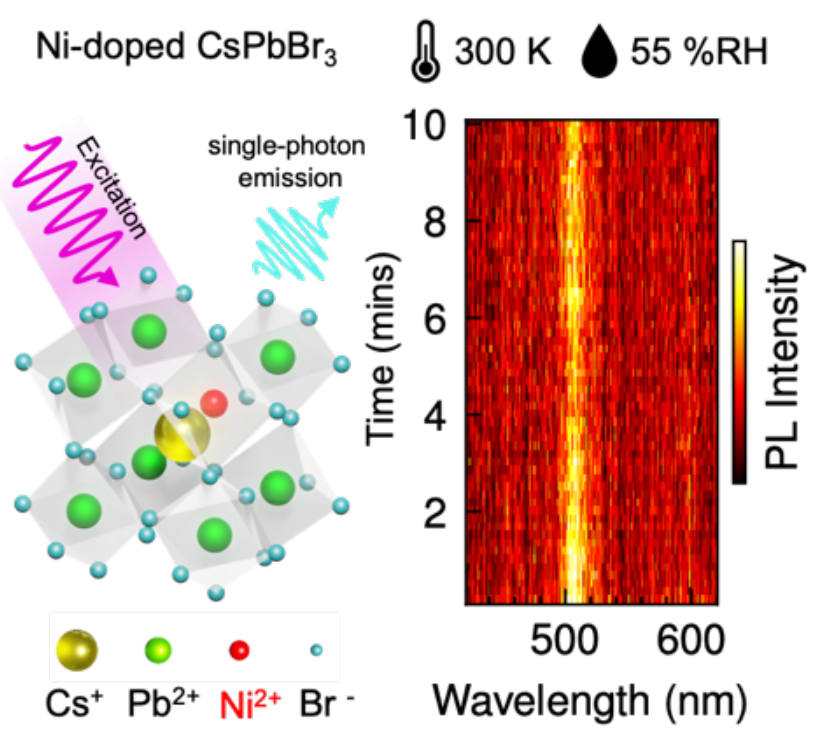